\documentclass[onecolumn,prc,superscriptaddress,unsortedaddress,showpacs,preprintnumbers,amsmath,amssymb]{revtex4}
\usepackage{tipa}

\usepackage{graphicx}
\usepackage{dcolumn}
\usepackage{bm}
\usepackage{color}

\def\beq{\begin{equation}}
\def\eeq{\end{equation}}
\def\bea{\begin{eqnarray}}
\def\eea{\end{eqnarray}}

\def\fun#1#2{\lower3.6pt\vbox{\baselineskip0pt\lineskip.9pt
  \ialign{$\mathsurround=0pt#1\hfil##\hfil$\crcr#2\crcr\sim\crcr}}}

\preprint{}

\begin{document}

\title{Coexistence of isospin $I=0$ and $I=1$ pairings in asymmetric nuclear matter}

\author{Y. -J. Yan}
 \affiliation{Institute
of Modern Physics, Chinese Academy of Sciences, Lanzhou 730000,
China}\affiliation{School of Nuclear Science and Technology,
University of Chinese Academy of Sciences, Beijing 100049, China}
\affiliation{CAS Key Laboratory of High Precision Nuclear
Spectroscopy, Institute of Modern Physics, Chinese Academy of
Sciences, Lanzhou 730000, China}
\author{X. -L. Shang}\email[ ]{shangxinle@impcas.ac.cn}
 \affiliation{Institute
of Modern Physics, Chinese Academy of Sciences, Lanzhou 730000,
China}\affiliation{School of Nuclear Science and Technology,
University of Chinese Academy of Sciences, Beijing 100049, China}
\affiliation{CAS Key Laboratory of High Precision Nuclear
Spectroscopy, Institute of Modern Physics, Chinese Academy of
Sciences, Lanzhou 730000, China}
\author{J.-M. Dong}
 \affiliation{Institute
of Modern Physics, Chinese Academy of Sciences, Lanzhou 730000,
China}\affiliation{School of Nuclear Science and Technology,
University of Chinese Academy of Sciences, Beijing 100049,
China}\affiliation{CAS Key Laboratory of High Precision Nuclear
Spectroscopy, Institute of Modern Physics, Chinese Academy of
Sciences, Lanzhou 730000, China}

\author{W. Zuo}
\affiliation{Institute of Modern Physics, Chinese Academy of
Sciences, Lanzhou 730000, China}\affiliation{School of Nuclear
Science and Technology, University of Chinese Academy of Sciences,
Beijing 100049, China}\affiliation{CAS Key Laboratory of High
Precision Nuclear Spectroscopy, Institute of Modern Physics,
Chinese Academy of Sciences, Lanzhou 730000, China}

\begin{abstract}
The coexistence of neutron-neutron (n-n), proton-proton (p-p), and
neutron-proton (n-p) pairings is investigated by adopting an
effective density-dependent contact pairing potential. These three
types of pairings can coexist only if the n-p pairing is stronger
than the n-n and p-p pairings for isospin asymmetric nuclear
matter. In addition, the existence of n-n and p-p pairs might
enhance n-p pairings in asymmetric nuclear matter when the n-p
pairing strength is significantly stronger than the n-n and p-p
ones. Conversely, the n-p pairing is reduced by the n-n and p-p
pairs when the n-p pairing interaction approaches n-n and p-p
pairings.
\end{abstract}


\maketitle

\section{INTRODUCTION}
The importance of pairing correlation in nuclear systems was
realized very early \cite{bohr}. In finite nuclei, neutron-neutron
(n-n) and proton-proton (p-p) pairing effects are realized in
several nuclear properties such as deformation, moments of
inertia,
alignments, and mass systematics \cite{nnimp1,nnimp2,nnimp3}. 
In extended systems, nuclear pairing is expected to occur in the
dense matter inside the neutron stars \cite{ns1,ns2}. This pairing
is crucial for understanding various phenomena in neutron star
physics, from the cooling of new born stars \cite{ns3,ns4} to the
afterburst relaxation in X-ray transients \cite{ns5}, as well as
in the understanding of glitches \cite{ns6}. However, insufficient
attention is paid to the isospin-singlet pairing, i.e., the
neutron-proton (n-p) paring. Recently, it was suggested that the
isospin-singlet pairing is possibly crucial in understanding of
some nuclear structural issues, such as the Gamow-Teller
transition \cite{npimp1,npimp2}. In addition, considering the spin
and isospin degrees of freedom, the nuclear Cooper pairs contain
very interesting inner structures \cite{huang}.

It is well-known that pair correlations crucially depend on the
pairing near the Fermi surface. Because neutrons and protons share
the same Fermi energy in symmetric nuclear matter, n-n (p-p) pairs
compete intensely with n-p pairs . Generally, the most
energetically favored excludes the others. The investigations on
nuclear pairs almost focus on a single pairing structure, i.e.,
either the n-n (p-p) or n-p pair only
\cite{shen,nn1,nn2,sh2,sdbaldo,np1,np2,np3,np4,sh3}. Nevertheless,
coexistence may emerge in special cases such as in the case of
isospin asymmetric nuclear matter. In a neutron-rich system,
although the isospin-singlet n-p pairing may be favored, the
excess neutrons can as well form isospin-triplet n-n pairs
coexisting with the other, and they can influence each other.
Furthermore, the nuclei far from the beta-stability line, i.e. the
exotic nuclei, can be obtained from heavy-ion collisions (HIC),
which has been addressed as a laboratory for the dynamic evolution
of the superfluid state of nuclear matter \cite{lab1}. New aspects
of pairing could appear in these exotic nuclei with regard to
isospin asymmetries, one of which might be the interplay between
n-n and n-p pairings in the nuclei owing to the significant
overlap of proton and neutron orbits \cite{huang,lab2}.

In Ref. \cite{huang}, the coexistence of isospin $I=1$ and $I=0$
pairing are considered to study the inner phase structure and
phase transition at low density where the BCS-BEC crossover
occurs. The result obtained indicates that including the $I=1$
channel pairing significantly alters the phase structure and phase
transition properties. In nuclear matter, another concern is the
interplay between the $I=1$ and $I=0$ pairings. Based on this
motivation, to investigate the coexistence of the n-n, p-p, and
n-p pairing in asymmetric nuclear matter with effective contact
pairing interaction in this study, we employ the extend
Nambu-Gorkov propagator, which includes the isospin triplet n-n
and p-p pairings  and the isospin singlet n-p pairing.

The paper is organized as follows: In Sec. II, we briefly derive
the gap equation and thermodynamics, as well as introduce the
adopted effective pairing interaction. The numerical results and
discussion are presented in Sec. III, where the results of the
coexistence of three types of pairings are compared with the
single pairing at certain density. Finally, a summary and a
conclusion are given in Sec. IV.

\section{Formalism}
The Nambu-Gorkov propagator at finite temperatures, including the
n-n, n-p, and p-p pairings \cite{huang}, is expressed as:
\begin{eqnarray}
G=\left(
\begin{array}{llll}
i\omega_{\upsilon}-\varepsilon_{n} & \ \ \ \  0 &\ \
\Delta_{np} &\ \    \Delta_{nn}\\ \\
\ \ \ \  0 & i\omega_{\upsilon}-\varepsilon_{p} & \ \
\Delta_{pp} &    -\Delta_{np}\\ \\
\ \   \Delta_{np} &\ \  \Delta_{pp} &
i\omega_{\upsilon}+\varepsilon_{p} &\ \ \ \  0 \\ \\
\ \  \Delta_{nn} &  -\Delta_{np} &\ \ \ \  0&
i\omega_{\upsilon}+\varepsilon_{n}
\end{array}
\right)^{-1},
\end{eqnarray}
where $\omega_{\upsilon}=(2\upsilon+1)\pi k_{B}T$ with
$\upsilon\in \mathbb{Z}$ represents the Matsubara  frequencies.
$\varepsilon_{n/p}=\textbf{p}^{2}/(2m)-\mu_{n/p}$ is the single
particle (s.p.) energy with chemical potential $\mu_{n/p}$. In
addition, $\Delta_{nn}$, $\Delta_{pp}$, and $\Delta_{np}$ are the
isospin-triplet n-n, isospin-triplet p-p, and isospin-singlet n-p
pairing gaps, respectively.

\subsection{Gap equations}
The neutron-proton anomalous propagator, which corresponds to
$G_{13}$, reads
\begin{small}
\begin{eqnarray}
&&F_{np}^{\dagger}(\omega_{\upsilon},\textbf{p})\nonumber\\&=&\frac{-\Delta_{np}
[(i\omega_{\upsilon})^{2}+i\omega_{\upsilon}(\varepsilon_{n}-\varepsilon_{p})-\varepsilon_{n}\varepsilon_{p}-\Delta_{np}^{2}-\Delta_{nn}\Delta_{pp}]}
{\big[(i\omega_{\nu})^{2}-E_{+}^{2}
\big]\big[(i\omega_{\nu})^{2}-E_{-}^{2} \big]}\nonumber\\
&=&\frac{
-\Delta_{np}\big\{[(i\omega_{\upsilon})^{2}-\varepsilon_{+}^{2}]-i\omega_{\upsilon}(2\delta\mu)+2\delta\mu^{2}+\frac{(\Delta_{nn}-\Delta_{pp})^{2}}{2}\big\}}
{\big[(i\omega_{\nu})^{2}-E_{+}^{2}
\big]\big[(i\omega_{\nu})^{2}-E_{-}^{2} \big]},\nonumber\\
\end{eqnarray}
\end{small}where
$E_{\pm}=\sqrt{\varepsilon_{+}^{2}\pm\sqrt{\varepsilon_{-}^{4}+\varepsilon_{\Delta}^{4}}}$
is the quasi-particle energy in the condensate with the definition
$\varepsilon_{\Delta}^{4}=\Delta_{np}^{2}[(\varepsilon_{n}-\varepsilon_{p})^{2}+(\Delta_{nn}-\Delta_{pp})^{2}]$
and
$2\varepsilon_{\pm}^{2}=\varepsilon_{n}^{2}+\Delta_{nn}^{2}+\Delta_{np}^{2}\pm(\varepsilon_{p}^{2}+\Delta_{pp}^{2}+\Delta_{np}^{2})$.
$\delta\mu=(\varepsilon_{p}-\varepsilon_{n})/2=(\mu_{n}-\mu_{p})/2$
represents the Fermi surface mismatch. The summation over the
Matsubara frequencies provides the density matrix of particles in
the condensate, i.e, the n-p pairing probabilities,
\begin{small}
\begin{eqnarray}
\nu_{np}(\textbf{p})&=&-\frac{\Delta_{np}}{2}\Big\{\big[\frac{1-2f(E_{+})}{2E_{+}}+\frac{1-2f(E_{-})}{2E_{-}}\big]\nonumber\\
&+&\frac{2\delta\mu^{2}+\frac{(\Delta_{nn}-\Delta_{pp})^{2}}{2}}{\sqrt{\varepsilon_{-}^{4}+\varepsilon_{\Delta}^{4}}}\big[\frac{1-2f(E_{+})}{2E_{+}}-\frac{1-2f(E_{-})}{2E_{-}}\big]\Big\}.\nonumber\\
\end{eqnarray}
\end{small}Here $f(E)=[1+\exp(\frac{E}{k_{B}T})]^{-1}$ is the well-known Fermi-Dirac
distribution function under a temperature $T$. Accordingly, the
n-p gap equation is expressed as
\begin{small}
\begin{eqnarray}
\Delta_{np}&=&\int\frac{d\textbf{p}}{(2\pi)^{3}}V_{np}\frac{\Delta_{np}}{2}\Big\{\big[\frac{1-2f(E_{+})}{2E_{+}}+\frac{1-2f(E_{-})}{2E_{-}}\big]\nonumber\\
&+&\frac{2\delta\mu^{2}+\frac{(\Delta_{nn}-\Delta_{pp})^{2}}{2}}{\sqrt{\varepsilon_{-}^{4}+\varepsilon_{\Delta}^{4}}}\big[\frac{1-2f(E_{+})}{2E_{+}}-\frac{1-2f(E_{-})}{2E_{-}}\big]\Big\}.\nonumber\\
\end{eqnarray}
\end{small}In the absence of the
n-n and p-p pairings, the quasi-particle energy $E_{\pm}$ becomes
$E_{\pm}=\sqrt{[(\varepsilon_{n}+\varepsilon_{p})/2]^{2}+\Delta_{np}^{2}}\pm\delta\mu=E_{\Delta}\pm\delta\mu$,
and the gap equation is reduced to a more familiar form for the
n-p pairing in asymmetric nuclear matter:
\begin{eqnarray}
\Delta_{np}&=&\int\frac{d\textbf{p}}{(2\pi)^{3}}V_{np}\frac{\Delta_{np}[1-f(E_{+})-f(E_{-})]}{2E_{\Delta}}.\nonumber\\
\end{eqnarray}
Similarly, the n-n and p-p pairing gaps are respectively expressed
as
\begin{small}
\begin{eqnarray}
\Delta_{nn}&=&\int\frac{d\textbf{p}}{(2\pi)^{3}}V_{nn}\frac{\Delta_{nn}}{2}\Big\{\big[\frac{1-2f(E_{+})}{2E_{+}}+\frac{1-2f(E_{-})}{2E_{-}}\big]\nonumber\\
&+&\frac{\varepsilon_{-}^{2}+\Delta_{np}^{2}(1-\frac{\Delta_{pp}}{\Delta_{nn}})}{\sqrt{\varepsilon_{-}^{4}+\varepsilon_{\Delta}^{4}}}\big[\frac{1-2f(E_{+})}{2E_{+}}-\frac{1-2f(E_{-})}{2E_{-}}\big]\Big\},\nonumber\\
\end{eqnarray}
\end{small}and
\begin{small}
\begin{eqnarray}
\Delta_{pp}&=&\int\frac{d\textbf{p}}{(2\pi)^{3}}V_{pp}\frac{\Delta_{pp}}{2}\Big\{\big[\frac{1-2f(E_{+})}{2E_{+}}+\frac{1-2f(E_{-})}{2E_{-}}\big]\nonumber\\
&-&\frac{\varepsilon_{-}^{2}+\Delta_{np}^{2}(\frac{\Delta_{nn}}{\Delta_{pp}}-1)}{\sqrt{\varepsilon_{-}^{4}+\varepsilon_{\Delta}^{4}}}\big[\frac{1-2f(E_{+})}{2E_{+}}-\frac{1-2f(E_{-})}{2E_{-}}\big]\Big\},\nonumber\\
\end{eqnarray}
\end{small}

The occupation numbers, corresponding to the matrix elements
$G_{11}$ and $G_{22}$, can be calculated by
\begin{small}
\begin{eqnarray}
n_{n}&=&\frac{1}{2}-\frac{\varepsilon_{n}}{2}\big[\frac{1-2f(E_{+})}{2E_{+}}+\frac{1-2f(E_{-})}{2E_{-}}\big]\nonumber\\
&-&\frac{\varepsilon_{-}^{2}\varepsilon_{n}-2\delta\mu\Delta_{np}^{2}}{2\sqrt{\varepsilon_{-}^{4}+\varepsilon_{\Delta}^{4}}}\big[\frac{1-2f(E_{+})}{2E_{+}}-\frac{1-2f(E_{-})}{2E_{-}}\big]
\nonumber\\
\end{eqnarray}
\end{small}and
\begin{small}
\begin{eqnarray}
n_{p}&=&\frac{1}{2}-\frac{\varepsilon_{p}}{2}\big[\frac{1-2f(E_{+})}{2E_{+}}+\frac{1-2f(E_{-})}{2E_{-}}\big]\nonumber\\
&+&\frac{\varepsilon_{-}^{2}\varepsilon_{p}-2\delta\mu\Delta_{np}^{2}}{2\sqrt{\varepsilon_{-}^{4}+\varepsilon_{\Delta}^{4}}}\big[\frac{1-2f(E_{+})}{2E_{+}}-\frac{1-2f(E_{-})}{2E_{-}}\big]
\nonumber\\
\end{eqnarray}
\end{small} The neutron and proton densities are respectively defined as
\begin{eqnarray}
\rho_{n}=2\int\frac{d\textbf{p}}{(2\pi)^{3}}n_{n}, \ \
\rho_{p}=2\int\frac{d\textbf{p}}{(2\pi)^{3}}n_{p}.
\end{eqnarray}Notably, the n-n, p-p, and
n-p pairing gaps couple to each other. For asymmetric nuclear
matter at the fixed neutron and proton densities, these gap
equations (4), (6), and (7) should be solved self-consistently
with the densities (10) at give densities and temperatures.

\subsection{Pairing interaction}
In principle, the nucleon-nucleon pairing interaction in nuclear
matter originates from the attractive component of the bare
two-body potential and the three-body force, and this pairing
interaction is modified by the nuclear medium, such as the
polarization effect \cite{sup2,sup3,sup4,sup5,scr1,scr2,ulbd}. In
this research, to obtain qualitative conclusions from the
coexistence of n-n, p-p, and n-p pairs, we adopt the
density-dependent contact interaction developed by Gorrido et al.
\cite{ddci} to model the pairing potential. For uniform nuclear
matter, the potential takes the form
\begin{eqnarray}
V_{I}(\textbf{r},\textbf{r}')=g_{I}\delta(\textbf{r}-\textbf{r}'),
\end{eqnarray}
with the effective coupling constant
\begin{eqnarray}
g_{I}=v_{I}[1-\eta_{I}(\rho_{I}/\rho_{0})^{\gamma_{I}}].
\end{eqnarray}
Here, $v_{I}$, $\eta_{I}$, and $\gamma_{I}$ are adjustable
parameters and $I=0,1$ denote the total isospin of the pairs. For
the n-n (p-p) pairing, $\rho_{I}=\rho_{n}$ ($\rho_{I}=\rho_{p}$)
and for the n-p pairing, $\rho_{I}=\rho_{n}+\rho_{p}$.
$\rho_{0}=0.17 \text{fm}^{-3}$ represents the saturation density.
Taking suitable values of the parameters, the pairing gap
$\Delta(k_{F})$ can be reproduced as a function of the Fermi
momentum $k_{F}=(3\pi^{2}\rho_{I})^{1/3}$ in the channel $L=0$,
$I=1$, $S=0$ (n-n and p-p) and $k_{F}=(3\pi^{2}\rho_{I}/2)^{1/3}$
in channel $L=0$, $I=0$, $S=1$ (n-p). We would like to emphasize
that there is also a kind of n-p pairing in the channel $L=0$,
$I=1$, $S=0$ for the symmetric nuclear matter. In this channel,
the n-p pairing force is approximately the same as the n-n or p-p
pairing force. As will be discussed in Sec. III, even a minor
asymmetry will destroy the n-p pairing in this channel. Therefore,
the $I=1$ pairings only represent neutron-neutron and
proton-proton pairings hereafter.

\begin{figure*}
\includegraphics[width=15 cm]{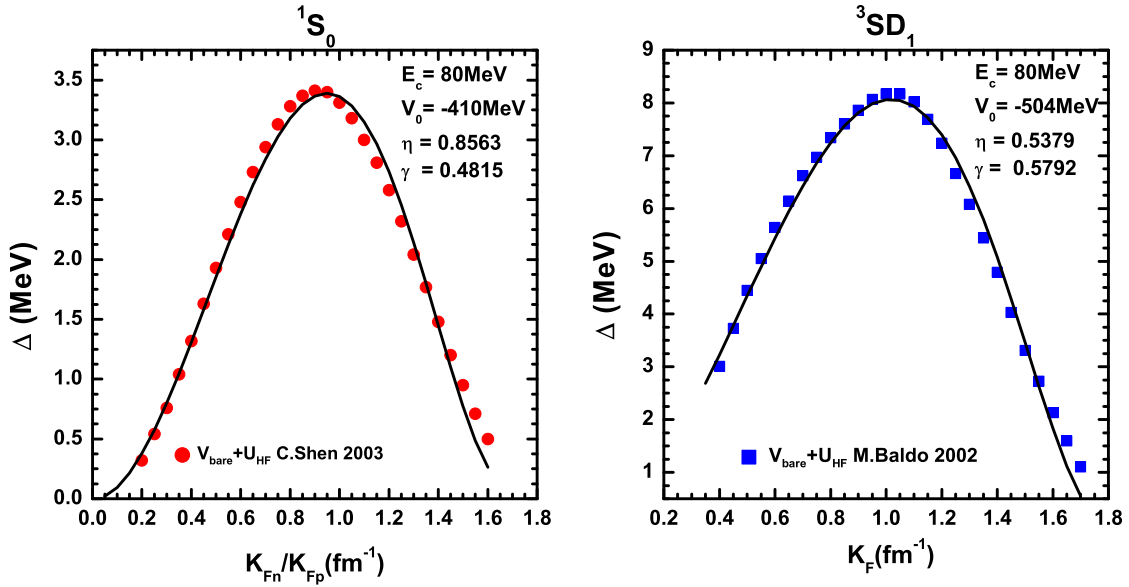}
\caption{\label{cshu} The density-dependent contact pairing
interaction with parameters calibrated to the calculated pairing
gaps. The dots represent the pairing gaps in Ref.
\cite{shen,sdbaldo}, whereas the lines correspond to the
calculation from the effective pairing interaction. The left
(right) panel is relate to the isospin triplet (singlet) channel.}
\end{figure*}
In addition to the polarization effect, the self-energy effect of
the medium quenches the pairing gaps \cite{shen,sh2}. Because the
self-energy effect for nuclear pairing remais an open question in
asymmetric nuclear matter, we adopt the calculated pairing gaps
\cite{shen,sdbaldo} under the Hartree-Fock approaches to calibrate
the parameters presented in Fig.1. It should be noted that the
self-energy \cite{sh2} and polarization \cite{ulbd} effects should
be included to obtain a more reliable pairing interaction. As is
well known that, to avoid the ultraviolet divergence, an energy
cut is required for the contact interaction. Here, we fix the
energy at approximately $80$ MeV for both cases. The left (right)
panel corresponds to the $I=1$ ($I=0$) pairings.

\subsection{Thermodynamics}
Now, we are in a position to determine the key thermodynamic
quantities. Because the occupation of the quasi-particle states is
given by the Fermi-Dirac distribution function, the entropy of the
system is obtained from
\begin{eqnarray}
S=-2k_{B}\sum_{\textbf{p}}\sum_{i}\big[f(E_{i})ln
f(E_{i})+\overline{f}(E_{i})ln \overline{f}(E_{i})\big],
\end{eqnarray}
where $\overline{f}(E_{i})=1-f(E_{i})$ and $i=\pm$. The internal
energy of the superfluid state is expressed as
\begin{small}
\begin{eqnarray}
U=2\sum_{\textbf{p}}\big[\varepsilon_{n}n^{n}+\varepsilon_{p}n^{p}\big]+\sum_{\textbf{p}}\big[g_{nn}\nu_{nn}^{2}+g_{pp}\nu_{pp}^{2}+2g_{np}\nu_{np}^{2}\big],\nonumber\\
\end{eqnarray}
\end{small}The factor $2$ corresponds to the spin summation.
The first term of Eq. (14) includes the kinetic energy of the
quasi-particle, as a function of the pairing gap and chemical
potential. The BCS mean-field interaction among the particles in
the condensate is embodied in the second term of Eq. (14). It
should be noted that for asymmetric nuclear matter, the n-n and
p-p pairing interactions can be different, i.e., $g_{nn}\neq
g_{pp}$, owing to $\rho_{n}\neq\rho_{p}$. Accordingly, the
thermodynamic potential can be given as
\begin{eqnarray}
\Omega=U-TS.
\end{eqnarray}

Once the contact pairing interaction is adopted, the pairing gap
is momentum independent. Therefore, the thermodynamic potential
can be obtained in a simple form:
\begin{small}
\begin{eqnarray}
\Omega&=&2\frac{\Delta_{np}^{2}}{g_{np}}+\frac{\Delta_{nn}^{2}}{g_{nn}}+\frac{\Delta_{pp}^{2}}{g_{pp}}
+\int\frac{d\textbf{p}}{(2\pi)^{3}}\nonumber\\
&\times&\Big\{\varepsilon_{n}+\varepsilon_{p}
-\sum_{i=\pm}\big[E_{i}+2k_{B}Tln(1+e^{\frac{-E_{i}}{k_{B}T}})\big]\Big\}.
\end{eqnarray}
\end{small}Her, We Consider the property $f(\omega)ln
f(\omega)+\overline{f}(\omega)ln\overline{f}(\omega)=-\frac{\omega}{k_{B}T}-ln(1+e^{-\omega/(k_{B}T)})$.
The gap equations (4), (6), and (7) and the densities of Eq. (10)
can be equivalently expressed as
\begin{small}
\begin{eqnarray}
\frac{\partial\Omega}{\partial\Delta_{np}}&=&0,\ \
\frac{\partial\Omega}{\partial\Delta_{nn}}=0,\ \
\frac{\partial\Omega}{\partial\Delta_{pp}}=0,\nonumber\\
\rho_{n}&=&-\frac{\partial\Omega}{\partial\mu_{n}},\ \ \ \
\rho_{p}=-\frac{\partial\Omega}{\partial\mu_{p}}.
\end{eqnarray}
\end{small}It should be noted that the solution of these equations
corresponds to the global minimum of the free energy
$F=\Omega+\mu_{n}\rho_{n}+\mu_{p}\rho_{p}$, which is the essential
quantity that describes the thermodynamics of asymmetric nuclear
matter.

\section{RESULTS AND DISCUSSION}

The numerical calculations in this study focus on the coexistence
of three different types of pairs in isospin asymmetric nuclear
matter with total density $\rho=\rho_{n}+\rho_{p}$ and isospin
asymmetry $\beta=(\rho_{n}-\rho_{p})/\rho$. We adopt the effective
contact pairing interaction at zero temperature. Fig.2 illustrates
the pairing gaps as a function of asymmetry $\beta$ at the total
density $\rho=0.068 \text{fm}^{-3}$, at which both the $I=1$ and
$I=0$ pairing interactions are most attractive. The thick lines
correspond to the results of the coexistence of three types of
pairings, which include $\Delta_{nn}\neq0, \Delta_{np}\neq0,
\Delta_{pp}\neq0$. In the symmetric matter, neutrons and protons
share the same Fermi surface, i.e., $k_{Fn}=k_{Fp}=k_{F}$,  and
the region near the Fermi surface contributes dominantly to the
pairing gaps. Two neutrons and two protons near the Fermi surface
can form a n-n pair and a p-p pair or two n-p pairs. Because the
n-p pairing strength is significantly stronger than that of n-n
and p-p, the nucleons prefer to form n-p pair instead of n-n (p-p)
pair. Equivalently, the n-p pairings severely suppress the n-n and
p-p pairings for $\beta=0$. As illustraed in Fig.2, the n-n (p-p)
gap disappears in symmetric case. In asymmetric nuclear matter,
the dominant region, which contributes significantly to the n-n
(p-p) pairing gap, is located at the neutron (proton) Fermi
momentum $k_{Fn}$ ($k_{Fp}$), whereas the region for n-p pairing
is between $k_{Fp}$ and $k_{Fn}$ (the average Fermi surface
related to the average chemical potential of neutrons and
protons). The split between neutron and proton Fermi surfaces
separates the dominant regions for n-n, p-p, and n-p pairings,
which enables the n-n and p-p pairing. And this discrepancy
between $k_{Fp}$ and $k_{Fn}$ increases with the increasing
isospin asymmetry. Therefore the n-n and p-p pairing gaps increase
with $\beta$.
\begin{figure*}
\includegraphics[width=15 cm]{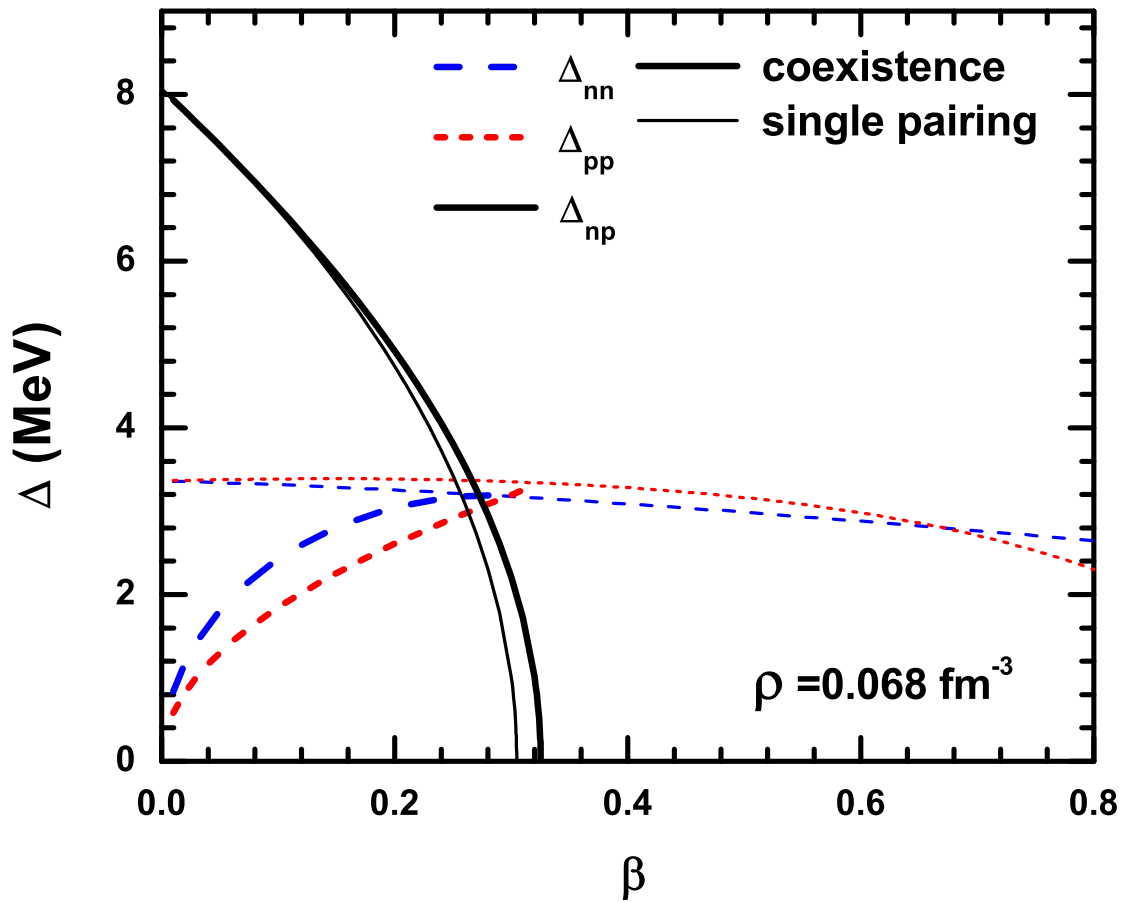}
\caption{\label{total} (Color online) The n-n, p-p, n-p pairing
gaps as a function of the isospin asymmetry $\beta$, at the total
density $\rho=0.068 \text{fm}^{-3}$. The thick and thin lines
correspond to the coexistence of three types of pairings and
single pairings, respectively. The dashed, short-dashed, and solid
lines are related to the n-n, p-p, and n-p pairings,
respectively.}
\end{figure*}

In addition, the results for single pairing, i.e.,
$\Delta_{nn}\neq0, \Delta_{pp}=\Delta_{np}=0$, $\Delta_{np}\neq0,
\Delta_{nn}=\Delta_{pp}=0, $, or $\Delta_{pp}\neq0,
\Delta_{nn}=\Delta_{np}=0, $ are depicted as thin lines in Fig. 2
for comparison. Owing to the suppression from the mismatched Fermi
surfaces, n-p pairing gaps decrease with $\beta$ and disappear at
certain asymmetries for both the single pairing and the
coexistence of three types of pairings. In the calculation of the
coexistence of three types of pairings, $\Delta_{nn}$ and
$\Delta_{pp}$ coincide with the results obtained from the single
pairing calculation when the n-p pairing vanishes. In fact, if
$\Delta_{np}=0$ the coupled equations (17) degenerates into two
groups of completely independent equations, which are the gap
equation for $\Delta_{nn}$ with the neutron density and the gap
equation for $\Delta_{pp}$ with proton density.

Compared to single pairing, the critical isospin asymmetry, where
$\Delta_{np}$ vanishes, is enhanced by the existence of n-n and
p-p pairs, as demonstrated in Fig. 2. Unfortunately, this
conclusion cannot be considered as definite, as the effective
pairing interaction is simply obtained from the pairing gaps under
the Hartree-Fock approximation. In addition, the effective n-p
pairing interaction can be significantly reduced by the
nucleon-nucleon correlation beyond the Hartree-Fock approaches
\cite{sh2}. Owing to the complexity of the nuclear many-body
medium effects, the exact effective pairing interaction remais an
open problem. To eliminate the uncertainty of the effective
pairing strength, we adjust the effective neutron-proton pairing
interaction artificially to obtain the qualitative conclusion. The
results obtained are presented in Fig. 3. The solid and dashed
lines correspond to the results obtained from the coexistence of
the three types of pairings and the single pairing, respectively.
For the effective interaction obtained from Ref. \cite{sdbaldo},
$g_{np}/g_{nn}=1.3837$. If we reduce the n-p pairing strength
$g_{np}$, the enhancement of the n-p pairing from the existence of
the n-n (p-p) pairs is reduced. When $g_{np}/g_{nn}$ is under a
certain value, the existence of n-n (p-p) pairing might suppress
the n-p pairing eventually. An interesting property is that if
$g_{np}\simeq g_{nn}$, $\Delta_{np}$ decreases rapidly with
$\beta$. As mentioned in Sec. II (B), the channel $L=0$, $I=1$,
$S=0$ embodies n-n, p-p, and n-p pairings, and the pairing
interactions are approximately the same for the asymmetric case. A
negligible asymmetry can destroy the n-p pairing in the $L=0$,
$I=1$, $S=0$ channel. Therefore, in general, the $I=1$ pairing
solely refers to the n-n and p-p pairings.
\begin{figure*}
\includegraphics[width=15 cm]{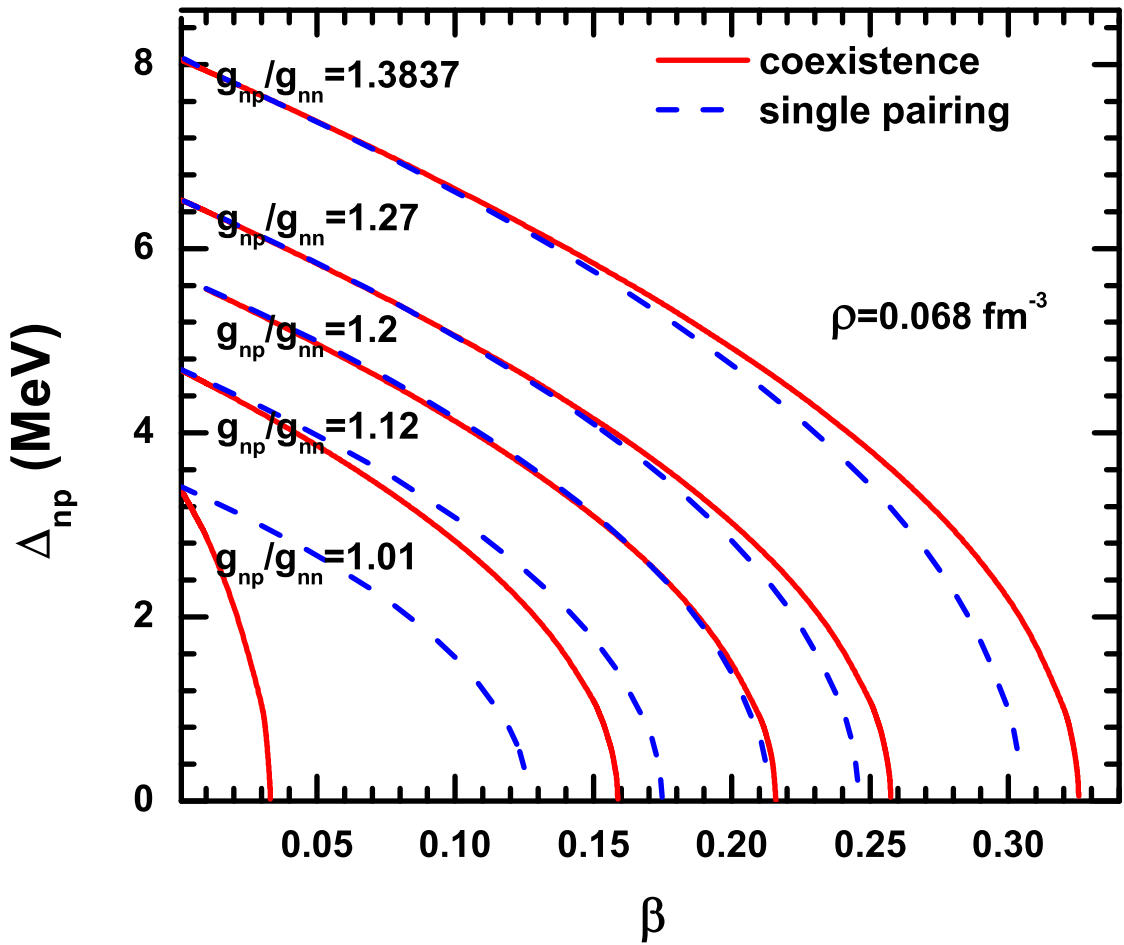}
\caption{\label{full} The n-p pairing gaps as a function of
isospin asymmetry at total density $\rho=0.068 \text{fm}^{-3}$ for
different n-p pairing strengths, $g_{np}/g_{nn}=1.3837$, $1.2$,
$1.12$, $1.01$. The solid and dashed lines correspond to the
coexistence of three types of pairings and single pairing,
respectively.}
\end{figure*}

One straightforward way to understand the enhancement of n-p
pairing from the existing n-n and p-p pairs is to investigate the
n-p pairing probabilities near the average Fermi surface (related
to the average chemical potentials of the neutron and proton). The
results obtained are depicted in Fig. 4, in the case where total
density $\rho=0.068 \text{fm}^{-3}$ and isospin asymmetry
$\beta=0.3$. The n-p pairing strength is set to be
$g_{np}/g_{nn}=1.3837$. For the single n-p pairing, the pairing is
forbidden in a window around the average Fermi surface owing to
the absence of protons. Once the n-n and p-p pairings are
included, the dispersion of neutron and proton Fermi surfaces can
provide the kinematical phase space near the average Fermi surface
for the occurence of the n-p pairing phenomena.  This is a
positive mechanism, such that the existence of n-n and p-p pairs
enhances the n-p pairing.
\begin{figure*}
\includegraphics[width=15 cm]{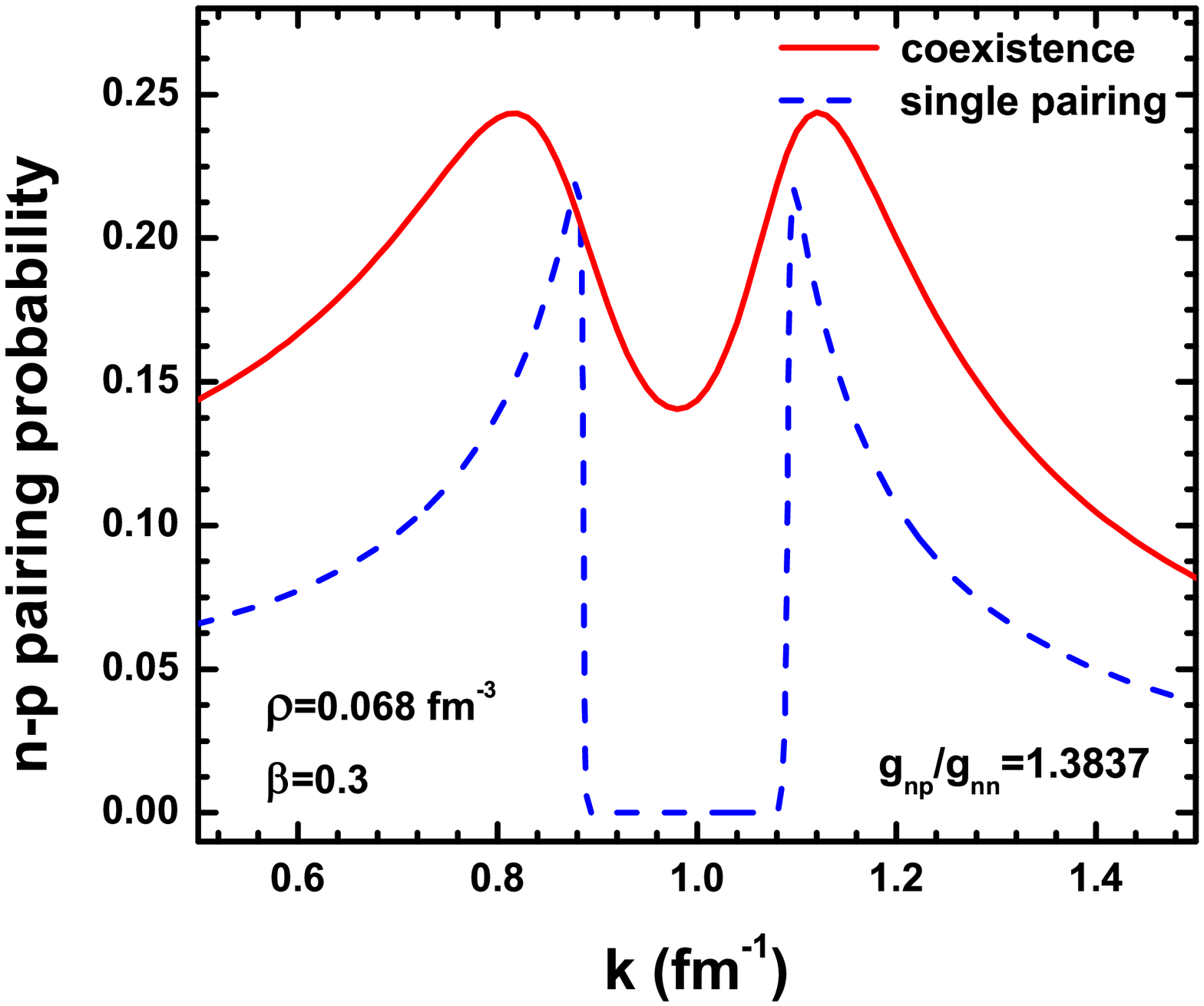}
\caption{\label{pp} The n-p pairing probabilities as a function of
$k$ near the average Fermi surface with the total density
$\rho=0.068 \text{fm}^{-3}$ and isospin asymmetry $\beta=0.3$.
Here, $k=p/\hbar$ is the wave number. The pairing strength is set
to be $g_{np}/g_{nn}=1.3837$. The solid and dashed lines
correspond to the coexistence of three types of pairings and
single pairing, respectively.}
\end{figure*}

Another effect of the existence of n-n and p-p pairs is that a n-n
pair and a p-p pair ought to be broken up to form two n-p pairs.
Exclusively, when the pairing energy of n-n and p-p pairs is
smaller than that of two n-p pairs, the existence of n-n and p-p
pairs can enhance the n-p pairing. The pairing energy is related
to the pairing strength directly. As presented in Fig. 5, when the
n-p pairing strength is insufficient, the n-p pairing probability
is suppressed significantly by n-n and p-p pairs.
\begin{figure*}
\includegraphics[width=15 cm]{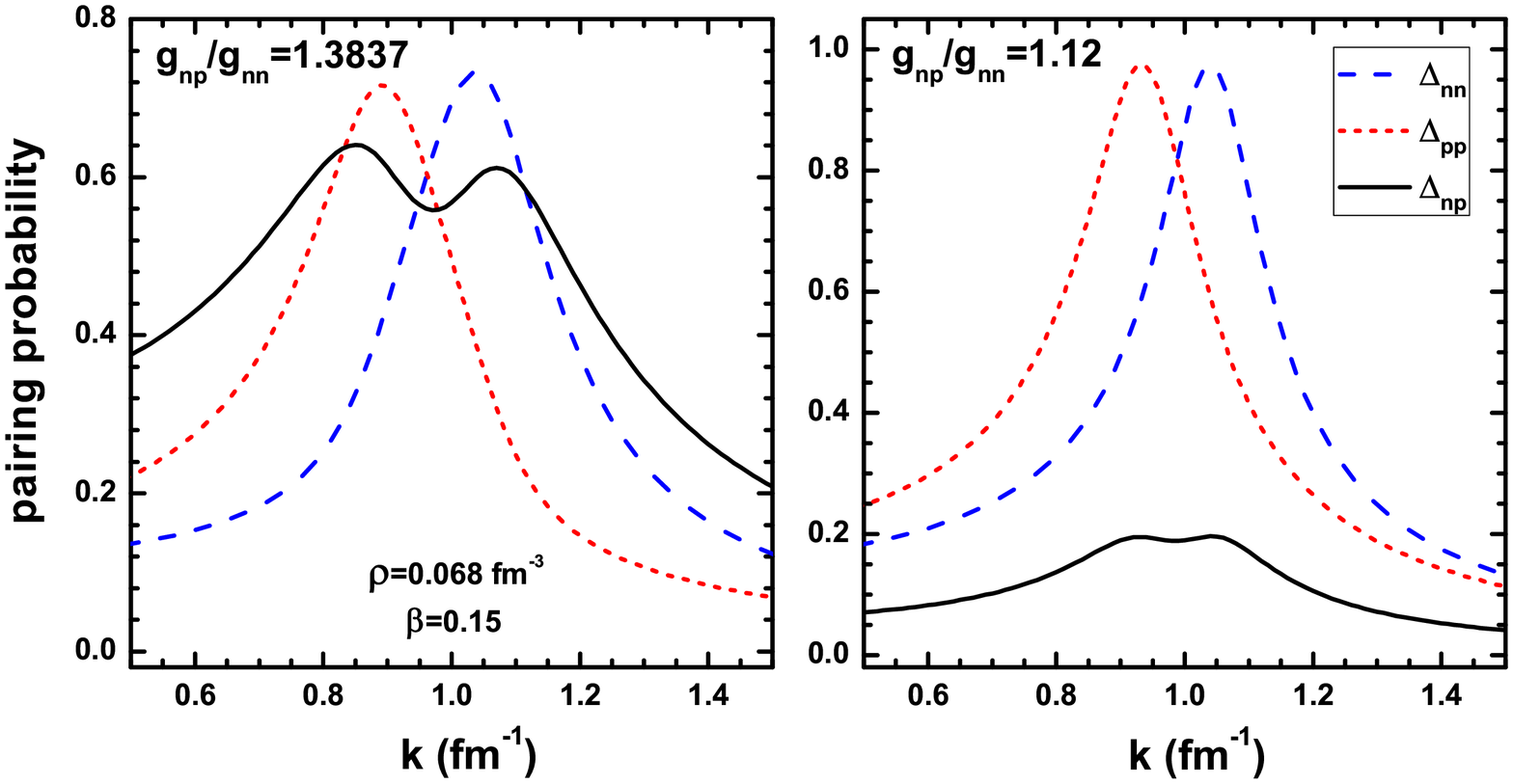}
\caption{\label{pp}The pairing probabilities vs $k$ near the
average Fermi surface at the total density $\rho=0.068
\text{fm}^{-3}$ with isospin asymmetry $\beta=0.15$. Here
$k=p/\hbar$ is the wave number. The dashed, short-dashed, and
solid lines correspond to the n-n, pp, and n-p pairings,
respectively. The n-p pairing strength $g_{np}/g_{nn}$ is set to
be $1.3837$ ($1.12$) in left (right) panel.}
\end{figure*}

In the calculations of this study, the temperature is set to be
zero. However, for asymmetric nuclear matter, the temperature can
also disperse the neutron and proton Fermi surfaces, which will
eventually reduce the suppression of Fermi surface mismatch at low
temperature. At high temperature, the temperature will destroy all
types of pairings. Once the temperature is included, the enhanced
and reduced effects on n-p paring from the existence of n-n and
p-p pairings should be weakened.

In finite nuclei, the n-p pairing might be suppressed by the
strong spin-orbit splitting \cite{sp1,sp2}. However, in nuclei
where the spin-splitting becomes small, the coexistence of three
types of pairings may occur. Understanding the enhanced and
reduced effects on n-p paring owing to the existence of n-n and
p-p pairings could be beneficial in elucidating the n-p pairing in
$N\approx Z$ nuclei. For asymmetric nuclei, the interplay between
n-n and n-p pairings might be the same as that in asymmetric
nuclear matter.

\section{SUMMARY}
In this study, we investigated the coexistence of n-n, p-p, and
n-p pairings in isospin asymmetric nuclear matter with an
effective density-dependent contact pairing interaction. The three
types of pairings cannot coexist in symmetric nuclear matter, only
n-p pairs can survive when the n-p pairing strength is stronger
than that of the n-n and p-p pairs, whereas the n-n and p-p pairs
are preferred if the n-n and p-p pairing interactions become
strong. Furthermore, n-n, p-p, and n-p pairs can coexist in
isospin asymmetric nuclear matter when the n-p pairing interaction
is stronger than n-n and p-p pairs.

Compared to the single pairing calculation (gap equation with only
one kind of nucleon pair), the results indicate two effects of the
existence of n-n and p-p pairs. On the one hand, the existence of
n-n and p-p pairs can disperse the neutron and proton Fermi
surfaces, which increase the phase-space overlap between neutrons
and protons and eventually enhance the n-p pairing near the
average Fermi surface. This positive mechanism can reduce the
suppression owing to the mismatched Fermi surface of neutrons and
protons in the isospin asymmetric nuclear matter. On the other
hand, a n-n pair and a p-p pair should be broken up to form two
n-p pairs. In this process, the pairing interaction plays a
crucial role. The final results are determined by these two
effects. In isospin asymmetric nuclear matter, the existence of
n-n and p-p pairs can enhance the n-p pairing when the n-p pairing
strength is significantly stronger than that of n-n and p-p pairs.
However, the existence of n-n and p-p pairs would reduce the n-p
pairing probability when the n-p pairing interaction decreases in
strength. Moreover, when the n-p pairing strength becomes
approximately that of n-n and p-p pairs, the n-p pairing rapidly
disappears with the isospin asymmetries.

In this paper, the gap solution is only thermodynamically stable.
The Cooper pair momentum should also be included in the future to
avoid dynamic instability \cite{ins1,ins2}. In addition, in future
works, the pairing interaction should be calibrated to the pairing
gaps, including the polarization correction and the correlation
effect. As a prospect, this interesting coexistence of the three
types of pairings would should also be applied to the studies on
pairing correlations in finite nuclei.

\section*{Acknowledgments}
{This work is supported by National Natural Science Foundation of
China (No. 11975282, 11775276, 11435014, 11505241), the Strategic
Priority Research Program of Chinese Academy of Sciences, Grant
No. XDB34000000, the Youth Innovation Promotion Association of
Chinese Academy of Sciences (Grant No. Y2021414, Y201871).}


\begin{thebibliography}{90}

\bibitem{bohr}
A. Bohr, B. R. Mottelson, and D. Pines, Phys. Rev. {\bf 100} 936
(1958).
\bibitem{nnimp1}
D. Brink, and R. Broglia, Nuclear Super uidity: Pairing in Finite
Systems (Cambridge University Press, Cambridge, 2005).
\bibitem{nnimp2}
R. A. Broglia, V. V. Zelevinsky (Eds.), 50 years of BCS, (World
Science Pub., 2012).
\bibitem{nnimp3}
D. J. Dean and M. Hjorth-Jensen, Rev. Mod. Phys. {\bf 75} 607
(2003).

\bibitem{ns1}
A. B. Migdal, Zh. Eksp. Teor. Fiz. {\bf 37} 249 (1959).
\bibitem{ns2}
G. Baym, C. Pethick, D. Pines, Nature {\bf 224} 673 (1969).

\bibitem{ns3}
J. M. Lattimer, K. A. Van Riper, M. Prakash, M. Prakash, ApJ {\bf
425}, 802 (1994).
\bibitem{ns4}
S. Burrello, M. Colonna, and F. Matera Phys. Rev. C {\bf 94},
012801(R) (2016).
\bibitem{ns5}
D. Page, S. Reddy, Neutron Star crust. edited by C. Bertulani and
J. Piekarewicz, Nova Science Publ., 281 (2012).
\bibitem{ns6}
J. Piekarewicz, F.J. Fattoyev, C.J. Horowitz, Phys. Rev. C {\bf
90}, 015803 (2014).
\bibitem{npimp1}
C. L. Bai, H. Sagawa, M. Sasano, T. Uesaka, K. Hagino, H.Q. Zhang,
X. Z. Zhang, and F.R. Xu, Phys. Lett. B {\bf 719} 116 (2013).
\bibitem{npimp2}
K. Kaneko, Y. Sun, and T. Mizusaki, Phys. Rev. C {\bf 97}, 054326
(2018).
\bibitem{huang}
S. J. Mao, X. G. Huang, and P. F. Zhuang, Phys. Rev. C {\bf 79},
034304 (2009).
\bibitem{shen}
Caiwan Shen, U. Lombardo, P. Schuck, W. Zuo, and N. Sandulescu,
Phys. Rev. C {\bf 67} 061302 (R) (2003).
\bibitem{nn1}
U. Lombardo, P. Schuck, and W. Zuo, Phys. Rev. C {\bf 64} 021301
(R) (2001).
\bibitem{nn2}
J. M. Dong, U. Lombardo, and W. Zuo, Phys. Rev. C {\bf 87} 062801
(R) (2013).
\bibitem{sh2}
X. -H. Fan, X. -L. Shang, J. -M. Dong, and W. Zuo, Phys. Rev. C
{\bf 99} 0665804 (2019).
\bibitem{sdbaldo}
M. Baldo, U. Lombardo, H. -J. Schulze, and Zuo Wei, Phys. Rev. C
{\bf 66} 054304 (2002).
\bibitem{np1}
U. Lombardo, H.-J. Schulze, and W. Zuo, Phys. Rev. C {\bf 59},
2927 (1999).
\bibitem{np2}
X. L. Shang, and W. Zuo, Phys. Rev. C {\bf 88}, 025806 (2013).
\bibitem{sh3}
X. L. Shang, P. Wang, P. Yin, and W. Zuo, J. Phys. G {\bf 42},
055105 (2015).
\bibitem{np3}
P. Bo\.{z}ek, Phys. Rev. C {\bf 62} 054316 (2000).
\bibitem{np4}
X. Meng, S. S. Zhang, L. Gio, L. S. Geng, and L. G. Cao, Phys.
Rev. C {\bf 102} 064322 (2020).
\bibitem{lab1}
M. Baldo, U. Lombardo and P. Schuck, Phys. Rev. C {\bf 52} 975
(1995).
\bibitem{lab2}
U. Lombardo, C. W. Shen, H. -J. Schulze, and W. Zuo, Int. J. Mod.
Phys. E {\bf 14} 513 (2005).
\bibitem{sup2}
J. Wambach, T.L. Ainsworth, D. Pines, Nucl. Phys. A {\bf 555} 128
(1993).
\bibitem{sup3}
H.-J. Schulze, J. Cugnon, A. Lejeune, M. Baldo, and U. Lombardo,
Phys. Lett. B {\bf 375} 1 (1996).
\bibitem{sup4}
H.-J. Schulze, A. Polls, and A. Ramos, Phys. Rev. C {\bf 63}
044310 (2001).
\bibitem{sup5}
C. W. Shen, U. Lombardo, and P. Schuck, Phys. Rev. C {\bf 71},
054301 (2005).
\bibitem{scr1}
L. G. Cao, U.Lombardo, P.Schuck, Phys. Rev. C. {\bf 74}, 064301
(2006).
\bibitem{scr2}
S. S. Zhang, L. G. Cao, U. Lombardo and P. Schuck, Phys. Rev. C
{\bf 93} 044329 (2016).
\bibitem{ulbd}
Wenmei Guo, U. lombardo and P. Schuck, Phys. Rev. C {\bf 99},
014310 (2019)
\bibitem{ddci}
E. Garrido \emph{et al.}, Phys. Rev. C {\bf 60}, 064312 (1999); 63
037304.
\bibitem{sp1}
G. F. Bertsch, and Y. Luo, Phys. Rev. C {\bf 81}, 064320 (2010).
\bibitem{sp2}
H. Sagawa, C. L. Bai, and G. Col\`{o}, Phys. Scr. {\bf 91}, 083011
(2016).
\bibitem{ins1}
I. M. Khalatnikov, Pis'ma Zh. Eksp. Teor. Fiz. {\bf 17}, 534
(1973) [JETP Lett. {\bf 17}, 386 (1973)]; V. P. Mineev, Zh. Eksp.
Teor. Fiz. {\bf 67}, 263 (1974) [Sov. Phys. JETP {\bf 40}, 132
(1974)]
\bibitem{ins2}
L. Y. He, M. Jin, and P. F. Zhuang, Phys. Rev. B {\bf 73} 214527
(2006).

\end{thebibliography}
\end{document}